\long\def\maintitle#1{{\vskip 1em
\begin{center}\section*{#1}\end{center}\nopagebreak[4]}}
\long\def\author#1{{\begin{center}\normalsize{\bf#1}\end{center}\vskip-1em
\index{#1}}\nopagebreak[4]}
\long\def\address#1{{\begin{center}\small\noindent#1\end{center}}\nopagebreak[4]}
\long\def\EAT#1\par#2\par#3\par#4\par%
{\newpage\vbox{%
\maintitle{#2}%
\author{#3}%
\address{#4}%
}\nopagebreak[4]}
\documentstyle[12pt]{article}

\begin{document}

\EAT

{Formation and migration of trans-Neptunian objects}

{S.~I.~Ipatov$^{1,2}$}

{$^1$Institute of Applied Mathematics, Moscow, Russia\\
$^2$NASA/GSFC, Greenbelt, USA}

\centerline{}

Abstracts of the Conference "Celestial Mechanics - 2002: Results and Prospects" (10 - 14 September 2002, St. Petersburg, Russia), Institute of Applied Astronomy of Russian Academy of Sciences.

\centerline{}

    We support the Eneev's idea [1] that the largest (with diameter $d$$>$100 km)
trans-Neptunian objects (TNOs) moving now in not very eccentric orbits could be formed
directly by the compression of rarefied dust condensations (with $a$$>$30 AU) of 
the protoplanetary cloud (we do not support Eneev's opinion that planets were 
formed directly from large condensations)
but not by the accretion of smaller (for example, 1-km) planetesimals, because
such accretion could take place only at a large total mass of TNOs (several tens of 
Earth masses) and very small eccentricities ($\sim 0.001$), which probably could not 
exist during the time needed for such accretion.
Probably, some planetesimals with $d\sim 100-1000$ km in the feeding zone of the giant 
planets and even large main-belt asteroids also could be formed directly by such 
compression. Some smaller objects (TNOs, planetesimals, and asteroids) could be 
debris of larger objects, and other such objects could be formed directly by 
compression of condensations. A small portion of planetesimals from the feeding zone 
of the giant planets that entered into the trans-Neptunian region could left in 
eccentrical orbits beyond Neptune and became so called ''scattered objects''. 
These objects could supply most of bodies that collided the Earth at 
the end of its bombardment 4 Gyr ago. 
The total amount of water delivered to the Earth during the formation of the giant 
planets was about the mass of water in the Earth's oceans. 

The motion of TNOs to Jupiter's orbit was investigated by several authors.
We considered the evolution for intervals $T_S$$\ge$5
Myr of 2500 Jupiter-crossing objects (JCOs) under the 
gravitational influence of all planets, except for Mercury 
and Pluto (without dissipative factors). In the first series 
we considered $N$=2000 orbits near the orbits of 30 real 
Jupiter-family comets with period $<$$10$ yr, and in the 
second series we took 500 orbits close to the orbit of Comet 
10P Tempel 2 ($a$$\approx$3.1 AU, $e$$\approx$0.53, $i$$\approx$$12^\circ$). 
We calculated the probabilities of collisions 
of objects with the terrestrial planets, using orbital 
elements obtained with a step equal to 500 yr and then 
summarized the results for all time intervals and all 
bodies, obtaining the total probability $P_\Sigma$ of 
collisions with a planet and the total time interval 
$T_\Sigma$ during which perihelion distance $q$ of bodies was 
less than a semimajor axis of the planet. The values of $P_r $$=$$10^6 P$$=$$ 
10^6 P_\Sigma /N$ and $T$$=$$T_\Sigma /N$ are presented 
in the Table together with the ratio $r$ of the total time 
interval when orbits were of Apollo type (at $a$$>$1 AU, $q$$=$$a(1-e)$$<$1.017 AU, $e$$<$$0.999$) 
to that of Amor type ($1.017$$<$$q$$<$1.33 AU); $r_2$ is the same as 
$r$ but for Apollo objects with $e$$<$0.9. 
For observed near-Earth objects $r$ is close to 1. 

\vspace{1mm}

 {\bf Table:} Values of $T$ (in kyr), $T_c$$=$$T/P$ (in Myr), $P_r$, $r$, 
$r_2$ for the terrestrial planets (Venus=V, Earth=E, Mars=M)

$ \begin{array}{llccccccccc} 

\hline

  & & $V$ & $V$ & $E$ & $E$ & $E$ & $M$ & $M$ & - & - \\

\cline{3-11}

 & N& T & P_r & T & P_r & T_c & T & P_r & r & r_2 \\

\hline

$JCOs$&2000&  9.3 & 6.62 & 14 & 6.65 &2110 & 24.7 & 2.03 & 1.32 &1.15\\
10P & 500& 24.9 & 16.3 & 44 & 24.5 & 1800 & 96.2 & 5.92 & 1.49 &1.34\\

\hline
\end{array} $ 

\vspace{1mm}

For integrations we used the Bulirsh-Stoer method (BULSTO) and a symplectic method. 
The probabilities of collisions of former JCOs with planets were close for these 
methods, but bodies got resonant orbits more often in the case of BULSTO. 
The obtained values of $T$ and $P_r$ are larger than those in [2], because in our last runs we 
considered much larger (than in [2]) number of JCOs and obtained several former JCOs that moved 
in orbits with aphelia inside Jupiter's orbit (mainly with $Q$$<$4.7 AU) during more than 1 Myr.
The probability of collisions with the Earth for 3 former JCOs from such orbits was 1.5 times greater than that for the other 1997 JCOs. 
About 1 of 300 JCOs collided with the Sun.
The analysis of the results of the orbital evolution of JCOs and TNOs showed that, 
in principle, as it was suggested earlier by T.M. Eneev, the trans-Neptunian belt can 
provide up to 100\% of Earth-crossing objects, 
but, of course, some of them came from the main asteroid belt. The ratio of the total 
mass of icy planetesimals that migrated from the feeding zone of the giant planets 
and collided with the planet to the mass of this planet was greater (by a factor of 
3 in our runs) for Mars than that for Earth and Venus. 

This work was supported by Russian Foundation for Basic Research~(01-02-17540), 
INTAS~(00-240), NASA~(NAG5-10776), and NRC~(0158730).

\end{document}